\documentclass{aa} 

\usepackage{graphicx}  
\usepackage{txfonts}  
          
\begin{document}  
  
\title{Asteroseismology of the \emph{Kepler} V777 Her variable white dwarf 
with fully evolutionary models}

\author{Alejandro H. C\'orsico\inst{1,2},  
        Leandro G. Althaus \inst{1,2},
        Marcelo M. Miller Bertolami \inst{1,2},\and
        Agn\`es Bischoff-Kim\inst{3}}
\institute{$^{1}$ Facultad de Ciencias Astron\'omicas y Geof\'{\i}sicas, 
           Universidad Nacional de La Plata, Paseo del Bosque s/n, 1900 
           La Plata, Argentina\\
           $^{2}$ CONICET\\
           $^{3}$ Chemistry, Physics and Astronomy Department, Georgia 
           College \& State University, Milledgeville, GA 31061, USA\\
           \email{acorsico,althaus,mmiller@fcaglp.unlp.edu.ar; 
           agnes.kim@gcsu.edu}     
           }

\date{\today}

\abstract{DBV stars  are pulsating white dwarfs  with atmospheres rich
  in  He.  Asteroseismology of  DBV stars  can provide  valuable clues
  about  the  origin, structure  and  evolution of  hydrogen-deficient
  white dwarfs,  and may  allow to study  neutrino and  axion physics.
  Recently, a  new DBV star, KIC  8626021, has been  discovered in the
  field of  the \emph{Kepler} spacecraft. It is  expected that further
  monitoring of this star in the next years will enable astronomers to
  determine  its   detailed  asteroseismic  profile.}{We   perform  an
  asteroseismological analysis  of KIC 8626021  on the basis  of fully
  evolutionary DB  white-dwarf models.}  {We employ a  complete set of
  evolutionary  DB white-dwarf  structures  covering a  wide range  of
  effective temperatures and stellar  masses.  They have been obtained
  on the basis of a  complete treatment of the evolutionary history of
  progenitors stars.  We  compute $g$-mode adiabatic pulsation periods
  for  this  set  of  models  and  compare  them  with  the  pulsation
  properties  exhibited by  KIC  8626021.}{On the  basis  of the  mean
  period spacing of the star, we found that the stellar mass should be
  substantially    larger   than    spectroscopy    indicates.    From
  period-to-period   fits  we   found  an   asteroseismological  model
  characterized  by  an effective  temperature  much  higher than  the
  spectroscopic     estimate.}{In    agreement    with     a    recent
  asteroseismological  analysis  of this  star  by  other authors,  we
  conclude that KIC  8626021 is located near the blue  edge of the DBV
  instability strip, contrarily  to spectroscopic predictions. We also
  conclude that the mass of KIC 8626021 should be substantially larger
  than thought.} 

  
\titlerunning{Asteroseismology of the DBV star KIC 8626021}
  
\authorrunning{C\'orsico et al.}  

\maketitle 

 
\section{Introduction}  
\label{intro}  

V777  Her (or  DBV)  stars  are $g$-mode  variable  white dwarfs  with
He-rich  atmospheres  (DB) and  effective  temperatures  in the  range
$21\,500 \lesssim  T_{\rm eff} \lesssim  29\,000$ K that  pulsate with
periods between 100  and $1100$ s (Winget \&  Kepler 2008; 
Althaus  et al. 2010).  They are  the hotter cousins of
the  ZZ Ceti  (or DAV)  stars, which  are pulsating  H-rich atmosphere
white dwarfs that define an instability strip centered at $T_{\rm eff}
\approx 12\,000$ K.  As an  excellent demonstration of the validity of
the  stellar pulsation  theory,  the  existence of  the  DBV class  of
compact pulsators was predicted  on theoretical grounds (Winget et al.
1982a)  before  it  were  confirmed  observationally  (Winget  et  al.
1982b).   Pulsations  in  V777 Her  are  thought  to  be driven  by  a
combination of the $\kappa-\gamma$  mechanism acting in the He partial
ionization  zone  ---and  thus  setting  the  blue  edge  of  the  DBV
instability  strip  (Winget et  al.   1983;  Bradley  \& Winget  1994;
C\'orsico  et  al. 2009),  and  the  ``convective driving''  mechanism
(Brickhill 1991; Goldreich \& Wu 1999) which is thought to be dominant
once the outer convection zone has deepened enough.

White-dwarf asteroseismology  ---the comparison between  the pulsation
periods  of white  dwarfs  and the  periods  computed for  appropriate
theoretical  models---  allows us  to  infer  details  of the  origin,
internal  structure and  evolution  of white  dwarfs.  In  particular,
estimates of the stellar mass,  He and H layer mass, core composition,
magnetic field, rotation  rate, seismological distance, and properties
of  the  outer convection  zone  can  be  inferred from  the  observed
pulsations of  DAV and DBV stars  (Winget \& Kepler  2008; 
Althaus et al. 2010).  Finally, an eventual measurement
of the temporal  changes in the observed stable  periods of DBVs could
allow  to  study neutrino  emission  (Winget  et  al. 2004)  and  also
\emph{axion} emission (see Isern et al. 2010 and references therein).

Unlike the ZZ Ceti variables,  that constitute the most numerous class
of pulsating white dwarfs (about 150 objects are known today;
Castanheira et  al. 2010),  V777 Her are  quite rare and  difficult to
find,  and until  recently, only  20 stars  of this  class  were known
(Beauchamp et al. 1999; Nitta et al. 2009; Kilkenny et al. 2009).  The
list   is  now   a  bit   enlarged   with  KIC   8626021  ($=$   GALEX
J192904.6$+$444708,  $T_{\rm eff}=  24\,950 \pm  750$ K  and  $\log g=
7.91\pm 0.07$ dex), a very recently discovered DBV star located in the
field  of view  of the  \emph{Kepler  Mission} by  {\O}stensen et  al.
(2011) (hereinafter  {\O}EA11)\footnote{Even more recently,  Hermes et
  al.  (2011) have reported the discovery of the first DAV star in the
  \emph{Kepler  Mission} field.}.   This star  exhibits at  least five
periodicities in  the range $197-376$  s and amplitudes between  1 and
5.2 mma,  with the three  strongest modes showing a  triplet structure
due possibly to  rotation.  It is expected that  additional modes with
higher radial orders (longer periods)  will be detected in future runs
of \emph{Kepler}, thus increasing the potential of asteroseismology to
infer its interior  structure.  Given the values of  $T_{\rm eff}$ and
$\log g$ quoted above and the DB models of Althaus et al. (2009a), KIC
8626021 should have a mass of $M_*= 0.56 \pm 0.03 M_{\odot}$.

Motivated by  the exciting discovery of the  first pulsating white
dwarf   in  the   \emph{Kepler}   field  of   view,   we  present   an
asteroseismological analysis of KIC 8626021 on the basis of the DB
white-dwarf models presented in Althaus et al.  (2009a).  This grid of
evolutionary models was computed for a wide range of stellar masses on
the  basis of  a complete  treatment  of the  evolutionary history  of
progenitors stars,  including those  stages relevant for  the chemical
profiles of the white dwarf, such  as the core H and He burning
phases, the  thermally pulsing asymptotic giant branch  phase, and the
born-again episode that is responsible for the H deficiency.

By considering the mean period spacing exhibited by the star, we found
that   KIC  8626021   should  be   substantially  more   massive  than
spectroscopy  indicates.  We  also found  that period  to  period fits
favour an  asteroseismological solution characterized  by an effective
temperature much higher than the spectroscopic estimate.

While   writing  this  paper,   Bischoff-Kim  \&   {\O}stensen  (2011)
(hereinafter BK{\O}11) announced the  results of their own independent
asteroseismological analysis on KIC 8626021. They found that this star
is actually a hot DBV.  Using entirely different models and methods, we
arrive at the same  conclusion. The fundamental difference between our
approach  and that  of BK{\O}11  is  that we  run evolutionary  models
starting  on  the  Zero  Age  Main Sequence,  accounting  for  nuclear
burning, time  dependent diffusion of the elements  and other physical
processes  that  take  place in  the  course  of  the evolution  of  a
star.  BK{\O}11 perform ``fast  white dwarf  asteroseismology'', where
they guess and parameterize the internal chemical composition profiles
and build static models. That  method allows them a fuller exploration
of parameter space. For KIC  8626021, they find very good fits (hotter
than  what the  preliminary spectroscopy  suggests) but  also conclude
that their internal Oxygen  composition profiles are in poor agreement
with stellar  evolution calculations. They suggest  further studies to
see what  the physical  parameters of models  that agree  with stellar
evolution calculations would be. In  essence, this is the kind of study
we present here. While our best fit models are evidently different, we
concur  with the  conclusion  that  KIC 8626021  is  hotter (and  more
massive)  than  suggested  by  the initial  spectroscopic  study.  The
conclusion that  this star  is residing  at the blue  edge of  the DBV
instability strip appears to be robust, and calls for the necessity of
a  new  and  improved  spectroscopic determination  of  its  effective
temperature.

In Section \ref{modeling} we present some details about our models and
methods.  In  Section  \ref{avgspacing},  we use  the  average  period
spacing  of KIC  8626021's period  spectrum to  draw  some conclusions
about its  mass and effective  temperature. We contrast our  models to
the grid of models computed by BK{\O}11. In Section \ref{best-fit}, we
present our best fit models. Again, we contrast our results with those
of BK{\O}11. In Section \ref{discussion} we discuss the discrepancies
between the results coming from spectroscopy and from 
asteroseismology, and the differences between our asteroseismological 
results and those of 
BK{\O}11. We conclude in Section \ref{conclusions}.

\section{Modeling}
\label{modeling}

\subsection{Numerical codes and asteroseismological approach}
\label{code}

\begin{figure}  
\centering  
\includegraphics[clip,width=250pt]{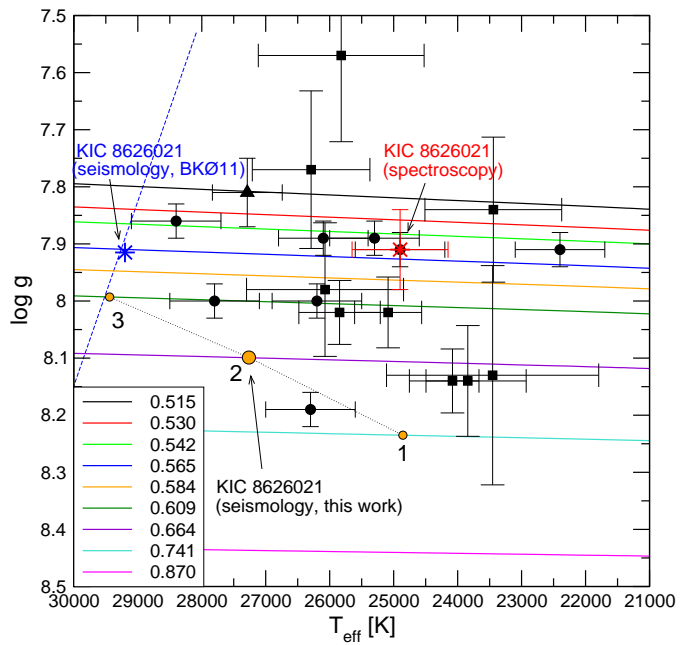}  
\caption{The location of the known DBV stars on the $T_{\rm eff}- \log
  g$ plane. Also  included are our DB white  dwarf evolutionary tracks
  displayed with  different colors according to the  stellar mass. The
  location of KIC 8626021 according to spectroscopy is emphasized with
  a  red star symbol,  while the  location of  our asteroseismological
  models for this star is  marked with orange circles connected with a
  dotted line.   The largest circle corresponds to  our best-fit model
  (model 2; see Sect.  \ref{best-fit}). For comparison, we include the
  location of the best-fit asteroseismological model of BK{\O}11 with a
  blue star symbol.  The theoretical  blue edge of the DBV instability
  strip corresponding to the version  MLT2 ($\alpha= 1.25$) of the MLT
  theory  of convection,  as derived  by C\'orsico  et al.  (2009), is
  depicted  with a  blue dashed  line.  Note:  the DBV  star  EC 05221
  (Kilkenny et al.  2009) is  not included because at present there is
  no any estimation of $T_{\rm eff}$ and $\log g$ available.}
\label{hr}  
\end{figure}  

The  evolutionary DB  white dwarf  models employed  in this  work were
presented in  Althaus et al.  (2009a) and we  refer to that  paper for
details.  Briefly,  the models were  computed with the  LPCODE stellar
evolutionary code we employed in our previous studies on the formation
of PG 1159  and extreme horizontal branch stars  (Althaus et al. 2005;
Miller Bertolami \&  Althaus 2006; Miller Bertolami et  al. 2008), hot
DQ white dwarfs  (Althaus et al.  2009b) as well  as the formation and
evolution of  He-core white  dwarfs with high  metallicity progenitors
(Althaus  et al.  2009c).   The LPCODE  evolutionary code  considers a
simultaneous  treatment  of non-instantaneous  mixing  and burning  of
elements,  which  is of  primary  importance  for  the calculation  of
chemical abundance changes  during the short-lived evolutionary stages
characteristic of unstable burning episodes, like the born-again stage,
from  which our  starting H-deficient  white dwarf  configurations are
derived. Nuclear reaction rates are from Caughlan \& Fowler (1988) and
Angulo et al.   (1999).  The $^{12}$C$(\alpha, \gamma)^{16}$O reaction
rate was  taken from Angulo  et al.  (1999),  which is about  twice as
large as  that of  Caughlan \& Fowler  (1988).  A  moderate, diffusive
overshooting  in  the core  and  in  the  envelope is  allowed  during
pre-white-dwarf evolution.  For the  white dwarf regime, we considered
the following main physical  ingredients.  Neutrino emission rates for
pair, photo,  and bremsstrahlung  processes are those  of Itoh  et al.
(1996).  For plasma  processes, we  use the treatment presented  in Haft et
al.   (1994).  Radiative  opacities  are  those of  the  OPAL  project
(Iglesias \&  Rogers 1996), including  C- and O-rich  composition.  We
adopted  the conductive  opacities  of Cassisi  et  al. (2007).   This
prescription  covers the  whole  regime where  electron conduction  is
relevant.  For the high density  regime, we used the equation of state
of  Segretain et  al.  (1994),  which accounts  for all  the important
contributions  for  both  the   liquid  and  solid  phases.   For  the
low-density  regime, we  used an  updated version  of the  equation of
state of  Magni \&  Mazzitelli (1979). Convection  was treated  in the
formalism   of  the  mixing-length   theory  as   given  by   the  ML2
parameterization  (Tassoul et  al.   1990).  All  of  our white  dwarf
sequences were computed in a  consistent way with the evolution of the
chemical abundance distribution caused  by element diffusion along the
whole  cooling  phase.   In  particular, we  considered  gravitational
settling  and  chemical  diffusion  of $^{4}$He,  $^{12}$C,  $^{13}$C,
$^{14}$N, and $^{16}$O.

The  adiabatic pulsation periods  employed in  the present  study were
assessed with the help of  the pulsational code described in C\'orsico
\& Althaus  (2006).  The prescription we  follow to assess  the run of
the  Brunt-V\"ais\"al\"a  frequency ($N$)  is  the so-called  ``Ledoux
Modified''  treatment ---see Tassoul  et al.   (1990)--- appropriately
generalized to include the effects of having three nuclear species (O,
C, and He) varying in abundance.

Our asteroseismological approach  basically consists in the employment
of detailed white dwarf models characterized by very accurate physical
ingredients.   These models  are  obtained by  computing the  complete
evolution of  the progenitor stars. We have  applied successfully this
approach to the hot DOVs or  GW Vir stars (see C\'orsico et al. 2007a,
2007b, 2008, 2009) and recently to an ensemble of bright ZZ Ceti stars
(Romero et al. 2011).  Since  the final chemical stratification of white
dwarfs is fixed  in prior stages of their  evolution, the evolutionary
history of progenitor stars is  of utmost importance in the context of
white dwarf asteroseismology.  Our asteroseismological approach, 
while being  physically  sound,  is by  far more  computationally 
demanding than  other approaches in which simplified  models are used.
As  a result,  our approach  severely  limits the  exploration of  the
parameter  space of  the models.   Indeed, for  the case  of  DB white
dwarfs, we  have only two  parameters which we  are able to vary  in a
consistent way: the stellar mass ($M_*$) and the effective temperature
($T_{\rm eff}$).   Instead, the  content of  He  ($M_{\rm He}$),  
the  shape  of the  C-O chemical structure  at the core 
(including the  precise proportions of central O and C), and the 
thickness of the chemical transition regions are fixed by the evolutionary 
history of progenitor stars.  

\begin{figure}
\centering
\includegraphics[clip,width=250pt]{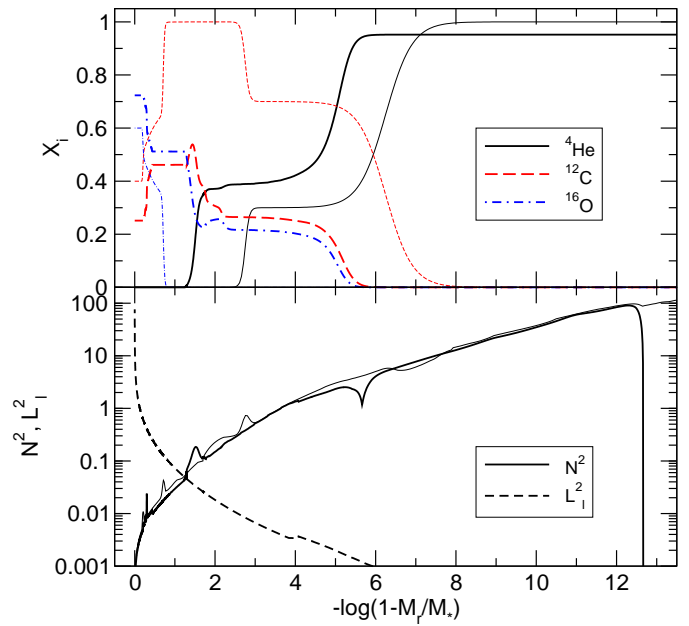}
\caption{The  internal  chemical  structure  (upper  panel),  and  the
  squared  Brunt-Va\"is\"al\"a  and  Lamb  frequencies for  $\ell=  1$
  (lower  panel) corresponding to  our template  DB white  dwarf model
  with a stellar mass $M_*= 0.565 M_{\odot}$, an effective temperature
  $T_{\rm  eff} \sim 28\,400$  K, and  a He  envelope mass  of $M_{\rm
    He}/M_* \sim 6.7 \times 10^{-3}$ (thick lines). For comparison, we 
    show with thin lines the chemical profiles and the Brunt-V\"ais\"al\"a
    frequency of the best fit model for KIC 8626021 found by BK\O11.}
\label{x-b-n2}
\end{figure}

\subsection{DB white dwarf evolutionary models}

The  initial models  for our  DB white  dwarf sequences  correspond to
realistic  PG  1159  stellar  configurations  derived  from  the  full
evolutionary calculations of  their progenitor stars (Miller Bertolami
\&  Althaus 2006).   All the  sequences  were computed  from the  ZAMS
through the  thermally-pulsating and mass-loss  phases on the  AGB and
finally to  the born-again  stage where the  remaining H  is violently
burnt.   After  the born  again  episode,  the H-deficient,  quiescent
He-burning  remnants evolve at  constant luminosity  to the  domain of
PG1159 stars with  a surface chemical composition rich in  He, C and O
(Miller Bertolami  \& Althaus 2006).   This new generation of  PG 1159
evolutionary  models has  succeed  in explaining  both  the spread  in
surface chemical  composition observed in  most PG 1159 stars  and the
location of  the GW Vir instability  strip in the 
$\log  T_{\rm eff}- \log g$  plane (C\'orsico  et al. 2006).   
Also, these PG  1159 models
have  been employed  in  detailed asteroseismological  studies of  six
pulsating  PG1159  stars  (C\'orsico  et al.   2007a,b,  C\'orsico  et
al. 2008, 2009).

Specifically, we considered nine DB white dwarf sequences with stellar
masses of:  $0.515, 0.530, 0.542, 0.565, 0.584,  0.609, 0.664, 0.741$,
and $0.870  M_{\odot}$.  These DB  sequences are characterized  by the
maximum He-rich  envelope that  can be left  by prior evolution  if we
assume that they are the result  of a born-again episode. The value of
envelope  mass ranges  from  $M_{\rm He}/M_*  \sim  2 \times  10^{-2}$
($M_*=  0.515 M_{\odot}$) to  $M_{\rm He}/M_*  \sim 1  \times 10^{-3}$
($M_*=  0.870  M_{\odot}$).   The  complete  set  of  DB  white  dwarf
evolutionary sequences  is displayed in Fig.   \ref{hr} with different
colors according to the value  of the stellar  mass.  Also shown  is the
location of the known DBV stars with values of $T_{\rm eff}$ and $\log
g$ extracted from  Beauchamp et al.  (1999) (for the  case in which no
traces of H are considered;  black circles), from Nitta et al.  (2009)
for the  nine DBV stars of  the SDSS (black squares),  and Kilkenny et
al. (2009) (black  triangle) .  The location of  KIC 8626021 according
to {\O}EA11 is highlighted with a red star symbol.

\subsection{A template model}

We  briefly describe  the main  pulsation properties  of our  DB white
dwarf  models.  We  focus  on a  template  DB model  with $M_*=  0.565
M_{\odot}$ and $T_{\rm eff} \sim  28\,400$ K.  In Fig. \ref{x-b-n2} we
depict the  internal chemical structure  of such model  (upper panel),
where  the  abundance  by  mass  of  the  main  constituents  ($^4$He,
$^{12}$C, and $^{16}$O)  is shown in terms of  the outer mass fraction
[$-\log(1-M_r/M_*)$].  The  chemical structure of  our models consists
of a C/O  core --- resulting from the core He  burning of the previous
evolution--- shaped  by processes  of extra mixing  like overshooting.
The core is surrounded  by a mantle rich in He, C,  and O which is the
remnant  of the  regions  altered by  the  nucleosynthesis during  the
thermally pulsing asymptotic giant branch.  Above this shell, there is
a pure He mantle with a mass $M_{\rm He}/M_* \sim 6.7 \times 10^{-3}$,
constructed by the action of  gravitational settling that causes He to
float to  the surface and  heavier species to sink.   A double-layered
structure  of  the He-rich  envelope  is  clearly  visible (Dehner  \&
Kawaler 1995;  Gautschy \& Althaus  2002; Althaus \&  C\'orsico 2004).
In our models, the shape  of the double-layered structure evolves with
time (with appreciable changes  even within the DBV instability strip)
by virtue of time-dependent  chemical diffusion processes (see Althaus
\& C\'orsico 2004).

The  lower panel of  Fig.  \ref{x-b-n2}  displays the  run of  the two
critical  frequencies of  nonradial stellar  pulsations, that  is, the
Brunt-V\"ais\"al\"a frequency and  the Lamb frequency ($L_{\ell}$) for
$\ell=  1$.  The  precise shape  of the  Brunt-V\"ais\"al\"a frequency
largely determines  the properties of the $g$-mode  period spectrum of
the  model.   In  particular,  each  chemical gradient  in  the  model
contributes locally  to the value  of $N$.  The most  notable features
are  the   very  peaked  feature   at  the  C/O   chemical  transition
[$-\log(1-M_r/M_*) \sim  0.3$], and the  bump at the  He/C/O interface
[$-\log(1-M_r/M_*)  \sim  1.5$].   On   the  other  hand,  the  He/C/O
transition region at $-\log(1-M_r/M_*) \sim 5$ is very smooth and does
not affect the pulsation spectrum much. For comparison, we also include 
in Fig. \ref{x-b-n2} the core chemical profiles and the Brunt-V\"ais\"al\"a 
frequency of the best fit model for KIC 8626021 found by BK{\O}11 
(thin lines).

\section{Period spacing: the global information}
\label{avgspacing}

For  $g$-modes  with  high   radial  order  $k$  (long  periods),  the
separation  of consecutive  periods ($|\Delta  k|= 1$)  becomes nearly
constant  at a  value  given  by the  asymptotic  theory of  nonradial
stellar  pulsations.   Specifically,  the  asymptotic  period  spacing
(Tassoul et al.  1990) is given by:

\begin{equation} 
\Delta \Pi_{\ell}^{\rm a}= \Pi_0 / \sqrt{\ell(\ell+1)},  
\label{aps}
\end{equation}

\noindent where

\begin{equation}
\label{asympeq}
\Pi_0= 2 \pi^2 \left[ \int_{r_1}^{r_2} \frac{N}{r} dr\right]^{-1}.
\end{equation}

\noindent This expression is rigorously valid for chemically homogeneous
stars.  In  principle, one can  compare the asymptotic  period spacing
computed from  a grid  of models with  different masses  and effective
temperatures with the  mean period spacing exhibited by  the star, and
then infer the value of the stellar mass. This method has been applied
in numerous  studies of  pulsating PG 1159  stars (see,  for instance,
C\'orsico et al. 2007a,b, 2008,  2009 and references therein). For the
method to be  valid, the periods exhibited by  the pulsating star must
be associated  with high  order $g$-modes, that  is, the star  must be
within  the asymptotic  regime  of pulsations.   KIC 8626021  exhibits
pulsation periods  associated to low  order modes, with  presumably $k
\lesssim 10$.  Furthermore, the interior  of DB white dwarf  stars are
supposed  to  be chemically  stratified  and  characterized by  strong
chemical gradients built up during  the progenitor star life.  So, the
direct  application of  the  asymptotic period  spacing  to infer  the
effective  temperature and  stellar  mass of  KIC  8626021 may  appear
somewhat  questionable.   However,  the  observed period  spectrum  is
roughly evenly spaced, indicating that using asymptotic theory may not
be an unreasonable thing to do.  In order to compare our models to the
period spacing  of the observed  pulsation spectrum, we  calculate the
average of the computed period spacings, using:

\begin{equation}
\label{avgdp}
\overline{\Delta  \Pi}_{\rm the}(M_*, T_{\rm eff})= \frac{1}{(n-1)} 
\sum_{k}^{n-1}  \Delta \Pi_k, 
\end{equation}

\noindent  where  $\Delta \Pi_k$  is  the  ``forward'' period  spacing
defined as $\Delta  \Pi_k= \Pi_{k+1}-\Pi_k$, and $n$ is  the number of
computed periods  laying in  the range of  the observed  periods.  The
theoretical period spacing  of the models as computed 
through Eq.  (\ref{aps}) and (\ref{avgdp})
share the same  general trends (that is, the  same dependence on $M_*$
and $T_{\rm  eff}$), although  $\Delta \Pi_{\ell}^{\rm a}$  is usually
somewhat higher  than $\overline{\Delta \Pi}_{\rm  the}$ provided that
this  last  quantity  is  computed  on a  range  of periods associated 
to low order
modes (short periods). We estimate  the observed mean period 
spacing  of KIC 8626021
by  means of  a nonlinear  least-squares fit  by considering  the five
periods observed in the star (first column of Table \ref{table2}).  We
obtain $\overline{\Delta \Pi}_{\rm obs}= 35.78 \pm 0.47$ s.

\begin{figure}
\centering
\includegraphics[clip,width=250pt]{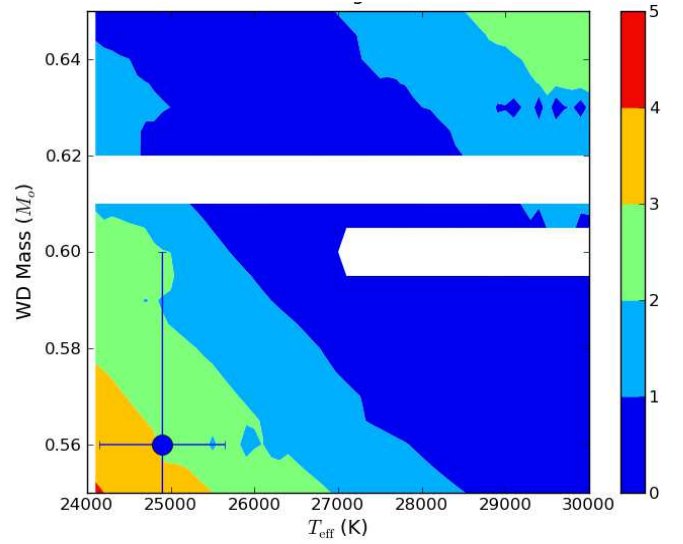}
\caption{
\label{avgdpgrid}
The location of the models  of BK{\O}11 with an average period spacing
close  to that  of KIC  8626021  in the  mass-temperature plane.   The
colour scale  is related  to the the  difference between  the observed
average period  spacing of  the star and  the period  spacing computed
with  Eq. (\ref{avgdp}).  The spectroscopic  values for  the  mass and
temperature of  the star are also  indicated with the  point and error
bars. It is  obviously outside the region of the  best fits in average
period spacing. The  white regions are regions where  models failed to
converge. They do not affect the results.}
\end{figure}

To check the validity of  using the period spacing to study
KIC 8626021, we  use the grid of models  computed by BK{\O}11 (400,000
models). For the fixed internal chemical composition profiles depicted
in Fig. \ref{x-b-n2}  and for each model, we  calculate the difference
between  the observed  average  period  spacing of  the  star and  the
calculated period  spacing [Eq. (\ref{avgdp})].  The result is  presented in
Fig. \ref{avgdpgrid}. A striking  feature is the diagonal region where
the  spacings are  close. This  is  a direct  consequence of  
Eq. (\ref{asympeq}) and evidence that  asymptotic theory applies at least in
broad  strokes.  In  Eq. (\ref{asympeq}), the  dependence  on  the
Brunt-V\"ais\"al\"a  frequency  is  such  that the  asymptotic  period
spacing is  greater when the mass  and/or temperature of  the model is
higher. This dependence is  apparent in Fig. \ref{avgdpgrid}. In order
to keep a constant average period  spacing (that of the star), one has
to compensate lower temperatures with higher masses and vice versa.

\begin{figure}  
\centering  
\includegraphics[clip,width=250pt]{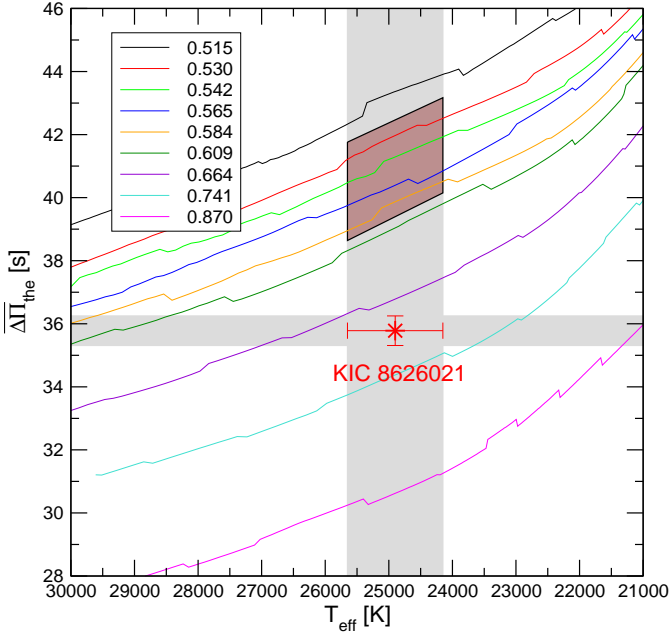}  
\caption{The average of the  computed period spacings corresponding to
  our  DB white  dwarf sequences  with different  stellar  masses. The
  location of  KIC 8626021  is shown with  a red star  symbol ($T_{\rm
    eff}= 24\, 900 \pm 750$ K, $\overline{\Delta \Pi}_{\rm obs}= 35.78
  \pm 0.47$ s). By simple linear interpolation, we found that the mass
  of the star according to its period spacing is 
  $M_*= 0.696\pm 0.031 M_{\sun}$.  According to spectroscopy 
  ($\log g= 7.91 \pm 0.07$; see Fig. \ref{hr}), the period
  spacing of the star should be somewhere within the dark shaded area.} 
\label{apsp}  
\end{figure}

In Fig.  \ref{apsp} we show the  run of average of the computed period
spacings [Eq. (\ref{avgdp})] in terms  of the effective temperature for all of
our DB  white-dwarf evolutionary  sequences.  The curves  displayed in
the  plot are  somewhat jagged.  This is  because the  average  of the
computed period spacings is evaluated for a fixed period interval, and
not  for a  fixed  $k$-interval.  As the  star  evolves towards  lower
effective temperatures, the periods  generally increase with time.  At
a given  $T_{\rm eff}$, there are  $n$ computed periods  laying in the
chosen period interval.   Later, when the model has  cooled enough, it
is  possible that  the  accumulated period  drift  nearly matches  the
period separation between adjacent  modes ($|\Delta k|= 1$).  In these
circumstances,  the number  of periods  laying in  the  chosen (fixed)
period interval  is $n \pm  1$, and $\overline{\Delta  \Pi}_{\rm the}$
exhibits  a  little  jump.   In  order  to  iron  out  the  curves  of
$\overline{\Delta \Pi}_{\rm the}$,  in constructing Fig. \ref{apsp} we
have  considered  pulsation  periods  in  a  wider  range  of  periods
($100\lesssim \Pi_k \lesssim  1200$ s) than that shown  by KIC 8626021
($195 \lesssim \Pi_k \lesssim 380$ s)
\footnote{If we adopt a shorter  range of periods, closer to the range
  of periods exhibited by KIC 8626021 (say $195-380$ s), the curves we
  obtain are much  more irregular and jumped, although  the results in
  relation to  KIC 8626021 do not appreciable  change.}.  The location
of KIC 8626021 is emphasized with  a red star symbol.  If we adopt the
spectroscopic effective temperature of KIC 8626021 ($T_{\rm eff}= 24\,
900 \pm  750$ K; {\O}EA11) we infer  a mass for the  star according to
its mean period  spacing of $M_*= 0.696\pm 0.031  M_{\odot}$, which is
$\sim 24  \%$ larger than  the spectroscopic estimate ($M_*=  0.56 \pm
0.03  M_{\odot}$).   The dark shaded area in the figure shows the 
location of the star ($\overline{\Delta \Pi}_{\rm the}, T_{\rm eff}$)
according to the constraint given on gravity by spectroscopy 
($\log g= 7.91 \pm 0.07$; see Fig. \ref{hr}). The strong discrepancy 
between the predictions of spectroscopy and the period spacing estimator 
is vividly depicted in this figure. This  difference  between  
the  spectroscopic  and
seismic masses is particularly relevant as, ultimately, both estimates
of the  stellar mass  of KIC 8626021  are based  on the same  DB white
dwarf evolutionary models.

This estimate of  the stellar mass is based on  the reliability of the
spectroscopic  determination of  the effective  temperature.   In this
sense,  it would  be  quite interesting  to  have an  estimate of  the
stellar mass devoid of any possible uncertainty from the spectroscopic
analysis.  If  we relax for the  moment the constraint  imposed by the
effective temperature, and  look for the possible range  of masses for
the star  to be within the  DBV instability strip  ($21\, 500 \lesssim
T_{\rm  eff}  \lesssim  29\  000$  K), we  find  that  $0.60  \lesssim
M_*/M_{\odot}  \lesssim 0.87$.   Confirming the  findings  of BK{\O}11
(see also  Fig. \ref{avgdpgrid}), the rather low  mass value predicted
by spectroscopy  ($M_* \sim 0.56  M_{\odot}$) is excluded by  the mean
period  spacing, unless  the  star is  outside  the instability  strip
($T_{\rm  eff}  \gtrsim  32\,000$  K;  see  Fig.   \ref{apsp}).   This
conclusion relies only on the measured mean period spacing, which is a
quantity  accurately measured,  and  on the  average  of the  computed
period spacing, a global quantity  that is not affected by the precise
shape  of  the chemical  transition  regions  of  the models.   It  is
particularly worth noting that this  result is independent of the mass
of the He-rich envelope. Indeed, Tassoul et al. (1990) have shown that
the  asymptotic period  spacing of  DBV white  dwarfs  is particularly
insensitive to the mass of  the He-rich mantle, changing less than $1$
s for a  change of 10 orders  of magnitude in the mass  of the He-rich
mantle (see their Figure 42).  Also, this conclusion is independent of
the effective temperature estimation of the star.  So, the result that
KIC  8626021   is  more  massive  and/or  hotter   than  indicated  by
spectroscopy  seems to be  a robust  finding\footnote{If we  apply the
  method described  above but employing the  asymptotic period spacing
  $\Delta \Pi_{\ell}^{\rm  a}$ [Eq. (\ref{aps})], we  obtain $M_* \sim
  0.80 M_{\odot}$, and if we relax the constraint of the spectroscopic
  $T_{\rm eff}$,  we find  that $0.66 \lesssim  M_*/M_{\odot} \lesssim
  0.90$ for the star to be within the DBV instability strip.}.

\section{Period fits: letting the pulsation periods to speak 
by themselves}
\label{best-fit}

Another  way to  infer the  stellar mass,  effective  temperature, and
details  of the  internal  structure  of KIC  8626021  is through  its
individual pulsation  periods, which  are extracted from  {\O}EA11 and
are  listed  in the  first  column  of  Table \ref{table2}.   In  this
approach we  seek a pulsation DB  white dwarf model  that best matches
the  pulsation periods  of KIC  8626021.  We  assume that  all  of the
observed periods correspond to $\ell=  1$ modes, to be consistent with
the  mean period  spacing.   The  goodness of  the  match between  the
theoretical  pulsation periods ($\Pi_k$)  and the  observed individual
periods  ($\Pi_{{\rm obs},  i}$) is  measured  by means  of a  quality
function defined as: 

\begin{equation}
\chi^2(M_*, T_{\rm eff})= \frac{1}{N} \sum_{i=1}^{N} 
\min[(\Pi_{{\rm obs},i}- \Pi_k)^2], 
\end{equation}

\noindent where $N$  (= 5) is the number of  observed periods.  The DB
white dwarf model  that shows the lowest value  of $\chi^2$ is adopted
as the ``best-fit model'' (see C\'orsico et al.  2007a,b, C\'orsico et
al.  2008,  2009, Romero  et  al.  2011).   We evaluate  the  function
$\chi^2(M_*, T_{\rm eff})$ for stellar masses of $0.515, 0.530, 0.542,
0.565, 0.584,  0.609, 0.664, 0.741$,  and $0.870 M_{\odot}$.   For the
effective  temperature we  employed a  much more  finer  grid ($\Delta
T_{\rm eff}= 10-30$ K).  The quality of our period fits is assessed by
means   of   the  average   of   the   absolute  period   differences,
$\overline{\delta}=  (\sum_{i=1}^N  |\delta_i|)/N$,  where  $\delta_i=
\Pi_{{\rm  obs}, i}  -\Pi_k$,  and by  the root-mean-square  residual,
$\sigma= \sqrt{(\sum |\delta_i|^2)/N}= \sqrt{\chi^2}$.

\begin{figure}  
\centering  
\includegraphics[clip,width=250pt]{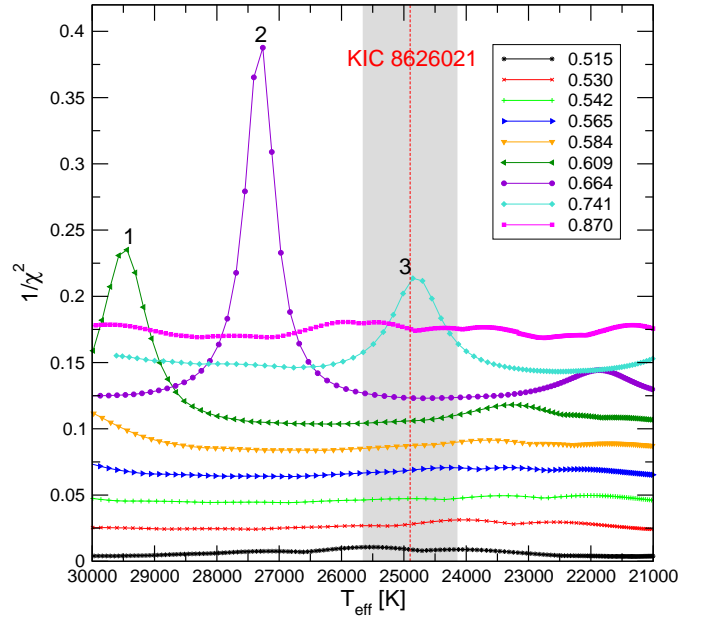}  
\caption{The  inverse of  the quality  function of  the period  fit in
  terms  of  the  effective  temperature.   The  vertical  gray  strip
  indicates  the  spectroscopic $T_{\rm  eff}$  and its  uncertainties
  ($T_{\rm  eff}=  24\,  900  \pm  750$  K).   The  curves  have  been
  arbitrarily  shifted  upward (with  a  step  of  0.02).  The  strong
  maximum in $(\chi^2)^{-1}$ corresponds  to the best-fit model (model
  2 in  Table \ref{table1}), with $M_*= 0.664\,  M_{\sun}$ and $T_{\rm
    eff}= 27\,263$ K.}
\label{chi2}  
\end{figure}

The quantity $(\chi^2)^{-1}$ in terms of the effective temperature for
different stellar masses is shown in Fig. \ref{chi2} together with the
spectroscopic effective temperature of  KIC 8626021 (red line) and its
uncertainties (gray strip).   We found one strong maximum  for a model
with $M_*= 0.664  M_{\odot}$ and $T_{\rm eff}= 27\,  260$ K (model 2).
Such  a pronounced  maximum  in  the inverse  of  $\chi^2$ implies  an
excellent   agreement    between   the   theoretical    and   observed
periods.  Notably, the  effective temperature  of this  model  is much
higher  (about  $2400$  K  larger) than  the  spectroscopic  effective
temperature  of  KIC  8626021.   Another  maximum,  albeit  much  less
pronounced, is encountered for  quite hotter and somewhat less massive
model with  $T_{\rm eff}= 29\,440$  and $M_*= 0.609  M_{\odot}$ (model
1).   According to  its  $T_{\rm  eff}$, this  model  is in  excellent
agreement  with  the asteroseismological  solution  found by  BK{\O}11
($T_{\rm eff}= 29\, 200$ K; see Table \ref{table3} below).  Finally, a
third  asteroseismological solution  is found  at a  model  with $M_*=
0.741 M_{\odot}$ and $T_{\rm eff}= 24\,856$ (model 3).  Interestingly,
this  model  has  exactly  the  spectroscopically  inferred  effective
temperature  of KIC  8626021, $T_{\rm  eff} \approx  24\,900 $  K.  In
Table \ref{table1} we summarize  the main characteristics of the three
solutions we found in our analysis, that is, models 1, 2 and 3. Models
1 and 3 constitute acceptable asteroseismological solutions.  However,
because  the agreement  between observed  and theoretical  periods for
these models are much poorer than for  model 2 (see columns 5 and 6 of
Table  \ref{table1}),  we  adopt  this  last  model  as  the  best-fit
asteroseismological model of KIC 8626021. One should bear in mind that
the analysis of  the mean period spacing suggests  a preferred mass of
$\sim  0.7  M_{\odot}$, between  our  0.664  $M_{\odot}$  and $  0.741
M_{\odot}$  models. Thus  a  better  fit, with  a  mass between  0.664
$M_{\odot}$ and $ 0.741 M_{\odot}$, might had arisen if a thinner mass
grid were available.

\begin{table}
\centering
\caption{Asteroseismological solutions for KIC  8626021.}
\begin{tabular}{ccccccc}
\hline
\hline
\noalign{\smallskip}
model &  $T_{\rm eff}$ & $\log g$ & $M_*$ & $\overline{\delta}$ &  $\sigma$ & BIC\\    
\noalign{\smallskip}
        & $[$K$]$       & $[$cm s$^{-2}]$  & $[M_{\odot}]$ & $[$s$]$ & $[$s$]$ &  \\
\noalign{\smallskip}
\hline
\hline
\noalign{\smallskip}
1 & $29\, 441$  &  7.993 & 0.609 & 2.314 & 2.721 & 1.15 \\
2 & $27\, 263$  &  8.099 & 0.664 & 1.582 & 1.934 & 0.85 \\
3 & $24\, 856$  &  8.235 & 0.741 & 3.334 & 3.685 & 1.41\\
\hline
\end{tabular}\\
\label{table1}
\end{table}

A detailed  comparison of the observed  $m= 0$ periods  in KIC 8626021
with the theoretical periods of the best-fit asteroseismological model
is  provided  in  Table  \ref{table2}.   For  this  model,  we  obtain
$\overline{\delta}= 1.582$  s and $\sigma=  1.934$ s. The  mean period
spacing of  our best  fit model is  $\overline{\Delta \Pi}=  35.10 \pm
0.50$ s (nonlinear least-squares fit), in excellent agreement with the
mean period spacing of  KIC 8626021 ($\overline{\Delta \Pi}_{\rm obs}=
35.78 \pm 0.47$).

BK{\O}11 perform  detailed period fits  to KIC 8626021  by considering
six parameters of  their DB white dwarf models:  $T_{\rm eff}$, $M_*$,
two parameters  describing the C-O core  composition profiles ($X_{\rm
  O}, q_{\rm fm}$)\footnote{BK{\O}11 define $X_{\rm O}$ as the central
  O  abundance and $q_{\rm  fm}$ as  the edge  of the  homogeneous C-O
  core.},  and  two  parameters  that define  the  envelope  structure
($M_{\rm env}, M_{\rm  He}$)\footnote{BK{\O}11 define $M_{\rm env}$ as
  the mass coordinate of the base of the He layer, and $M_{\rm He}$ as
  the mass  coordinate of the point  where the He  abundance raises to
  1.}.  They fix $M_{\rm env}$ and vary the remainder 5 parameters. In
contrast, in our models we  have only 2 free parameters: $T_{\rm eff}$
and $M_*$, and the chemical structure at the core and envelope is kept
fixed according  to the predictions of  the evolutionary computations.
In order  to compare the quality of  our best fit with  the results of
BK{\O}11,  we compute the  Bayes Information  Criterion (BIC;  Koen \&
Laney 2000):

\begin{equation}
{\rm BIC}= N_{\rm p} \left(\frac{\log N}{N} \right) + \log \sigma^2,
\end{equation}

\noindent where $N_{\rm p}$ is  the number of free parameters, and $N$
the number of  observed periods. In our case,  $N_{\rm p}= 2$ (stellar
mass and  effective temperature).  The  smaller the value of  BIC, the
better the quality  of the fit. We obtain ${\rm  BIC}= 0.85$, which is
substantially  larger than  the BIC  value of  the best  fit  model of
BK{\O}11  (${\rm BIC}=  -0.41$).  This  means that  our period  fit is
somewhat  poorer than  theirs. Notwithstanding  this, our
asteroseismological model  still provides  a very satisfactory  fit to
the periods of KIC 8626021.

\begin{table}
\centering
\caption{Comparison between the observed periods of KIC 8626021 
and the theoretical periods of the best-fit model (model 2 in Table \ref{table1}).}
\begin{tabular}{ccccccc}
\hline
\hline
\noalign{\smallskip}
$\Pi_{{\rm obs},i}$ &  $A_i$ &  $\Pi_k$ &  $\ell$ & $k$ & $\delta_i$ & $d\Pi_k/dt$\\    
\noalign{\smallskip}
$[$s$]$        & $[$ mma$]$ &  $[$s$]$ & & & $[$s$]$ & $[10^{-14}$ s/s$]$\\
\noalign{\smallskip}
\hline
\hline
\noalign{\smallskip}
197.11 & 2.92 &  200.43 & 1 & 3 & $-3.32$ & 5.60  \\ 
232.02 & 5.21 &  233.69 & 1 & 4 & $-1.67$ & 8.66  \\
271.60 & 1.87 &  271.22 & 1 & 5 & $ 0.38$ & 6.84  \\
303.56 & 1.29 &  301.42 & 1 & 6 & $ 2.14$ & 9.91  \\
---    & ---  &  339.69 & 1 & 7 &  ---    & 6.67  \\
376.10 & 1.05 &  376.50 & 1 & 8 & $-0.40$ & 12.13 \\
\hline
\end{tabular}\\
\label{table2}
\end{table}

\begin{table}
\centering
\caption{The main  characteristics of  KIC 8626021. The  second column
  corresponds to  spectroscopic results ({\O}EA11),  whereas the third
  and  fourth  columns present  results  from the  asteroseismological
  study  of BK{\O}11 and  from the  best-fit asteroseismological model  
  of this work, respectively.}
\begin{tabular}{lccc}
\hline
\hline
\noalign{\smallskip}
Quantity                    & {\O}EA11             & BK{\O}11     & This work  \\
\hline
\noalign{\smallskip}
$T_{\rm eff}$ [K]           & $24\,950 \pm 750$  &  $29\,200$     & $27\,263$  \\
$M_*$ [$M_{\odot}$]          & $0.56\pm 0.03$     &  $0.570$       & $0.664$    \\ 
$\log g$ [cm/s$^2$]         & $7.91 \pm 0.07$    &  ---           & $8.099$    \\ 
$\log (L_*/L_{\odot})$       & ---                &  ---           & $-1.14$    \\  
$\log(R_*/R_{\odot})$        & ---                &  ---           & $-1.93$    \\  
$M_{\rm He, total}$ [$M_{\odot}$] & ---             & ---            & $3.6 \times 10^{-3}$ \\  
\noalign{\smallskip}
\hline
\noalign{\smallskip}
$M_{\rm env}\ ^{(*)}$         & ---                & $-2.80$        & $-1.63$    \\
$M_{\rm He}\ ^{(*)}$          & ---                & $-6.30$        & $-5.95$   \\ 
$X_{\rm O}\ ^{(**)}$          & ---                & $0.60-0.65$    & $0.65$   \\
$q_{\rm fm}\ ^{(***)}$         & ---               & $0.36$         & $0.46$   \\
\noalign{\smallskip}
\hline
\hline
\end{tabular}
\label{table3}

{\footnotesize  Notes: 
$^{(*)}$ Mass coordinate, defined as $\log(1-M_r/M_*)$ (see  BK{\O}11).
$^{(**)}$ Central abundance by mass.
$^{(***)}$ Coordinate $M_r$ in units of $M_*$ (see  BK{\O}11).}

\end{table}

The last column in Table  \ref{table2} shows the rate of period change
of the  fitted pulsation  modes. Our calculations  predict all  of the
pulsation periods  to \emph{increase} with  time ($\dot{\Pi}_k>0$), in
accordance with  the decrease of the  Brunt-V\"ais\"al\"a frequency in
the core of  the model induced by cooling. Note  that at the effective
temperature  of  KIC  8626021,  cooling  has  the  largest  effect  on
$\dot{\Pi}_k$, while gravitational contraction, which should result in
a  \emph{decrease} of  periods with  time, becomes  negligible  and no
longer affects  the pulsation periods.   Until now, no  measurement of
$\dot{\Pi}$ in a DBV has  been assessed, although important efforts to
measure the  rate of period  change in at  least one star at  the blue
edge  (EC 20058$-$5234;  Dalessio et  al.  2010)  are  being currently
carried  out, and  a possible  determination of  the rate  of period
change for other DBV star (PG 1351+489) has been reported (Redaelli et
al. 2011).

The  main features  of  our  best-fit model  are  summarized in  Table
\ref{table3},  where we  also include  the parameters  of  KIC 8626021
extracted  from  {\O}EA11  and  the  seismological  model  derived  by
BK{\O}11.  In the Table,  the quantity $M_{\rm He, total}$ corresponds
to the total content of He of  the envelope of the model.  In order to
make the comparison easy, we  include in the table the four parameters
employed by BK{\O}11 that define the chemical profiles at the core and
envelope of their models. Note  that BK{\O}11 do not specify the value
of $M_{\rm  He, total}$ of their  best-fit model. The  location of the
best-fit model  for KIC  8626021 both according to our study and that 
of BK{\O}11 in the  $\log T_{\rm eff}-\log  g$ plane is shown in Fig.  
\ref{hr}. Both asteroseismic studies lead to the conclusion that 
KIC 8626021  should be closer to the blue edge  of the DBV instability
strip than spectroscopy suggests.

\section{Discussion}
\label{discussion}

The  results of  our asteroseismological  analysis point  to  a higher
stellar mass of KIC 8626021 than predicted by spectroscopy.  We arrive
at such conclusion through both  the period spacing and the individual
periods exhibited  by the star.  Regarding  the effective temperature,
our  work indicates  a  higher $T_{\rm  eff}$  than the  spectroscopic
measurement.

Our  results also  differ  somewhat from  those  of the  seismological
analysis of BK{\O}11.   Specifically, we obtain an asteroseismological
model that is more massive and  cooler than that of BK{\O}11. The fact
that in both independent analysis a good match to the observed periods
is found can be understood on  the basis of the asymptotic behavior of
$g$-mode pulsations, that predicts  that a lower effective temperature
is compensated by  a higher mass. What is  interesting and exciting is
that,  in spite  of the  substantial  differences in  the white  dwarf
modeling  (in  particular,  the quite different  
composition profiles,  that  lead  to significant  differences in  
the pulsation  periods)  both analysis
agree that KIC 8626021  is a hot DBV. This is in agreement with analyses 
based on the average period spacing and also with the fact that low  period modes are present in KIC 8626021's pulsation spectrum, as is also observed for the hot DBV EC20058.

\section{Conclusions}
\label{conclusions}

In  this paper  we have  presented  a detailed asteroseismic  
analysis of  KIC
8626021, the  first pulsating  DB white dwarf  star discovered  by the
\emph{Kepler  Mission},  on the  basis  of  the  full evolutionary  DB
white-dwarf  models presented in  Althaus et  al.  (2009a)  which were
computed for a  wide range of stellar masses  and He envelopes.  These
DB  white  dwarf  models  are  characterized  by  consistent  chemical
profiles for both the core  and the envelope.  These chemical profiles
are the result  of the computation of the  full and complete evolution
of the progenitor stars from the zero age main sequence, including the
core H and He  burning phases, the thermally pulsing asymptotic
giant branch phase, the born-again episode that is responsible for the
H  deficiency, and from  time-dependent element  diffusion predictions
during the white-dwarf stage.

By considering the mean period spacing exhibited by the star, we found
that  KIC  8626021 should  have  a stellar  mass  in  the range  $0.60
\lesssim M_*/M_{\odot} \lesssim 0.87$, substantially larger than those
derived  by   previous  spectroscopic  ($M_*   \sim  0.56  M_{\odot}$;
{\O}EA11)  and  asteroseismic ($M_*  \sim  0.57 M_{\odot}$;  BK{\O}11)
studies.  We  also  found  that  period-to-period  fits  point  to  an
asteroseismological  model  with  an  effective temperature  of  $\sim
27\,300$  K,  in  strong  conflict  with  the  spectroscopic  estimate
($T_{\rm eff}  \sim 24\, 900$ K).   Our results are  in agreement with
the  recent asteroseismic  analysis  of BK{\O}11  on  KIC 8626021,  in
particular  regarding  its  effective  temperature.   In  fact,  these
authors  conclude that  KIC  8626021  \emph{must be}  a  hot DBV  with
$T_{\rm eff} \sim 29\,200$ K. It would be interesting to see what a
spectroscopic analysis based on higher signal-to-noise spectra will tell
about the surface gravity and effective temperature of the star.

If KIC 8626021 is a hot  DBV, as first found by BK{\O}11 and confirmed
now  by our  results,  then  further monitoring  of  KIC 8626021  with
\emph{Kepler} in the  next years probably will allow  a measurement of
$\dot{\Pi}$, which in turn  could open the possibility to constrain the
plasmon neutrino emission rate (Winget et al. 2004; BK{\O}11).   
This endeavour is evidently dependent on the models, which can 
always be improved. Uncertainties in the models can also be 
assessed (Bischoff-Kim et. al. 2008).


\begin{acknowledgements}  
This work  was supported  by AGENCIA: Programa  de Modernizaci\'on
Tecnol\'ogica BID 1728/OC-AR, and by PIP 2008-00940 from CONICET.
This research  has made use of NASA's Astrophysics Data
System.
\end{acknowledgements}  
  

\end{document}